\documentclass[12pt]{article}

\usepackage{epsfig}

\def\bphi{\mbox{\boldmath $\phi$}}

\def\balpha{\mbox{\boldmath $\alpha$}}
\def\bbeta{\mbox{\boldmath $\beta$}}

\def\bb{\mbox{\bf b}}

\def\bt{\mbox{\bf t}}
\def\bc{\mbox{\bf c}}

\begin{document}
\begin{titlepage}

\baselineskip 24pt

\begin{center}

{\Large {\bf The Framed Standard Model (I) - A Physics Case for Framing the 
Yang-Mills Theory?}}*

\vspace{.5cm}

\baselineskip 14pt

{\large CHAN Hong-Mo}\\
h.m.chan\,@\,stfc.ac.uk \\
{\it Rutherford Appleton Laboratory,\\
  Chilton, Didcot, Oxon, OX11 0QX, United Kingdom}\\
\vspace{.2cm}
{\large TSOU Sheung Tsun}\\
tsou\,@\,maths.ox.ac.uk\\
{\it Mathematical Institute, University of Oxford,\\
Radcliffe Observatory Quarter, Woodstock Road, \\
Oxford, OX2 6GG, United Kingdom}

\end{center}
\vspace{.3cm}

\begin{abstract}

Introducing, in the underlying gauge theory of the Standard Model, the frame 
vectors in internal space as field variables (framons), in addition to the usual
gauge boson and matter fermions fields, one obtains: 
\begin{itemize}
\item the standard Higgs scalar as the framon in the electroweak sector; 
\item a global $\widetilde{su}(3)$ symmetry dual to colour to play the role of
fermion generations.
\end{itemize}
Renormalization via framon loops changes the orientation in generation space of 
the vacuum, hence also of the mass matrices of leptons and quarks, thus making 
them rotate with changing scale $\mu$.   From previous work, it is known already
that a rotatiing mass matrix will lead automatically to:
\begin{itemize}
\item CKM mixing and neutrino oscillations,
\item hierarachical masses for quarks and leptons,
\item a solution to the strong-CP problem transforming the theta-angle into a
Kobayashi-Maskawa phase.
\end{itemize}
Here in the FSM, the renormalization group equation has some special properties
which explain the main qualitative feaures seen in experiment both for mixing 
matrices of quarks and leptons, and for their mass spectrum.  Quantitative
results will be given in (II).  The paper ends with some tentative predictions
on Higgs decay, and with some speculations on the origin of dark matter.

\end{abstract}

*Invited talk at the Conference on 60 Years of Yang-Mills Gauge Thories,
IAS, Singapore, 25-28 May 2015.

\end{titlepage}

\clearpage

When the Yang-Mills theory was discovered 60 years ago, its significance was
immediately recognized, although it was unclear at that stage in what physics 
context it should be applied.  Now, 60 years later, the theory has found itself 
enthroned as the theoretical basis of, among other things, the standard model
of particle physics, namely of all known physical phenomena apart from gravity.
And this standard model can justly claim to be the most successful theory ever,
given the range of phenomena it covers and the resilience it has shown in
surviving the many detailed experimental tests to which it has been subjected.

However, some things are still missing from this beautiful picture, at least
to the fastidious theoretical mind, notably an understanding of the origin of:

\begin{itemize}

\item The Higgs boson needed to break the electroweak symmetry,

\item Three generations of quarks and leptons needed to fit experiment, 

\end{itemize}
neither of which is part of the original Yang-Mills structure or has any other
theoretical explanation.  The lack of understanding of the second, especially, 
is practically significant, since the masses and mixing matrices of quarks and 
leptons fall into a bizarre hierarchical pattern, and they account for about
two-thirds of the standard model's twenty-odd empirical parameters.  There are 
besides some ominous clouds on the horizon, such as the unresolved strong 
CP-problem, the missing right-handed neutrino, and the mysterious dominance of 
dark matter in the universe, plus, of course, the wanting link to gravity 
already mentioned.

To address these short-comings of the standard model, if indeed short-comings 
they are, one will obviously have to enrich its starting assumptions in some 
way.  One direction, the most popular one, is to keep the Yang-Mills structure
as it is, but enrich the superstructure by enlarging the gauge symmetry beyond 
the standard $su(3) \times su(2) \times u(1)$ (GUT, Supersymmetry), or the 
dimension of space-time beyond the standard 3 + 1 (Kaluza-Klein), or both 
(Superstring, branes). These extensions often open up grand vistas of things to 
come but are less successful in answering some of the detailed questions of 
immediate interest.  For example, instead of reducing the number of parameters 
in the standard model by explaining the known values of some of them, 
supersymmetric models usually end up with a hundred parameters or more.  For 
this reason, one thought, it might be worth trying a different path. 

Suppose one keeps the same gauge symmetry $su(3) \times su(2) \times u(1)$ and 
the same 4-dimensional space-time as for the standard model, but enrich instead 
the underlying structure by requiring the Yang-Mills theory be ``framed''.  By 
``framing'' a gauge theory here, we mean the introduction also as dynamical 
variables of the frame vectors in the internal symmetry space in addition to the
usual gauge potential $A_\mu$ and the fermionic matter fields $\psi$.  Such 
frame vector (or ``framon'') fields are analogues of the vierbeins $e_\mu^a$ in 
the theory of gravity in which they are often used in place of the metric 
$g_{\mu\nu}$ as dynamical variables.  By taking them as dynamical variables too 
in the particle theory makes it closer in spirit to the gravity theory, and may 
eventually facilitate the union of the two.  Our immediate aim, however, is not
to attempt this union but, while staying within particle physics itself, to 
address the shortcomings of the standard model listed above.

What then will framing give us in particle theory?  Like the vierbeins $e_\mu^a$
in gravity, the framons we need for the particle theory may be regarded as 
column vectors of a matrix relating the local gauge frame to a global reference 
frame, carrying hence both a local symmetry index (analogous to $\mu$ in 
$e_\mu^a$) and a global frame index (analogous to $a$ in $e_\mu^a$).  They thus 
transform both under the local gauge transmformations of $G = su(3) \times su(2)
\times u(1)$ and under the global transformations of the reference frame, say
$\tilde{G} = \widetilde{su}(3) \times \widetilde{su}(2) \times \tilde{u}(1)$,
but are just scalars under the Lorentz transformations in space-time.  Two
immediate consequences of framing then result: 

\begin{itemize}

\item {\bf (A)}  In the electroweak sector, the framon is an $su(2)$ doublet 
but Lorentz scalar field, exactly as needed for the Higgs field to break the 
electroweak symmetry;

\item {\bf (B)}  Since physics is independent of the choice of the reference 
frame, the framed theory has to be invariant not only under the local symmetry 
$G$, but also under its dual, the global symmetry $\tilde{G}$.  Of this the 1st 
component $\widetilde{su}(3)$, a 3-fold symmetry, can function as fermion 
generation, while the 2nd component $\widetilde{su}(2)$ can act as up-down 
flavour, and the 3rd component $\tilde{u}(1)$ as $(B-L)$,

\end{itemize}
giving thus both the Higgs field and fermion generation each an hitherto lacking
geometric significance.

To push these advantages further, there is some ambiguity first to settle as 
to the exact form that the framons should take.  Minimality considerations in 
the number of scalar fields to be introduced suggest then that the framons in 
FSM belong to the representation $({\bf 3} + {\bf 2}) \times {\bf 1}$ in $G$ but
to $\tilde{\bf 3} \times \tilde{\bf 2} \times \tilde{\bf 1}$ in $\tilde{G}$ and 
that some of its components may be taken as dependent on others \cite{efgt,dfsm}
leaving just:

\begin{itemize}
\item a ``weak framon'' of the form:
\begin{equation}
\balpha \otimes \bphi
\label{wframon}
\end{equation}
where $\balpha$ is a triplet in $\widetilde{su}(3)$, which may be taken without 
loss of generality \cite{dfsm} as a real unit vector in generation space, but
is constant in space-time, while $\bphi$ is an $su(2)$ doublet but Lorentz 
scalar field over space-time which has the same properties as, and may thus be 
identified with, the standard Higgs field;
\item the ``strong framon'':
\begin{equation}
\bbeta \otimes \bphi^{\tilde{a}},\quad  \tilde{a} = \tilde{1}, \tilde{2},
   \tilde{3}
\label{sframon}
\end{equation}
where $\bbeta$ is a doublet of unit length in $\widetilde{su}(2)$ space but 
constant in space-time, while $\bphi^{\tilde{a}}$ are 3 colour $su(3)$ triplet 
Lorentz scalar fields over space-time, which when taken as column vectors give 
a matrix $\Phi$ transforming by $su(3)$ transformation from the left but by 
$\widetilde{su}(3)$ transformations from the right.
\end{itemize} 

As usual the mass matrices at tree-level of quarks and leptons are to be 
obtained from the Yukawa couplings of the fermions to the Higgs scalar field 
(weak framon) by replacing it with its vacuum expectation value.  But now since 
the weak framon (\ref{wframon}) carries a factor $\balpha$, the fermion mass 
matrices will also carry this factor.  Then by a simple relabelling of the 
right-handed singlet fields, 

\begin{itemize}

\item {\bf (C)} The mass matrices of all quarks and leptons can be rewritten 
conveniently in the following form:
\begin{equation}
m = m_T \balpha^{\dagger} \balpha,
\label{mfact}
\end{equation}
where $\balpha$, coming from the framon,  is ``universal'', i.e. the same for 
up-type quark (U), down-type quark (D), charges leptons (L) and neutrinos (N), 
and only the coefficient $m_T$ depends on the fermions species. 

\end{itemize}
Now such a mass matrix has long been coverted by phenomenologists \cite{Fritsch,
Harari} as a starting approximation, since it gives only 1 massive generation 
for each species, which may be interpreted as embryonic mass hierarchy, and zero
mixing between up- and down-states, which is not a bad approximation, at least
for quarks.  

The question now, of course, is what happens above the tree-level.  This is the
point at which the FSM first shows its power, beyond what can be done by just
phenomenology.  Because of the double invariance under both $G = su(3) \times 
su(2) \times u(1)$ and $\tilde{G} = \widetilde{su}(3) \times \widetilde{su}(2) 
\times \tilde{u}(1)$, the action for framons is much restricted in form, which 
allows some radiative corrections to be calculated.  In particular, since the 
strong framon in (\ref{sframon}) above carries both the local colour $su(3)$ and
global dual colour $\widetilde{su}(3)$ (or generation) indices, renormalization 
by strong framon loops will change the orientation of the vacuum in generation 
space.  This change will depend in general on the renormalization scale $\mu$, 
thus inducing a $\mu$-dependent $\widetilde{su}(3)$ transformation (rotation) on
the vector $\balpha$ which appears in the fermion mass matrix (\ref{mfact}).

Now, we have studied the consequences of a rotating rank-one mass matrix (R2M2)
for some years and it has been shown, e.g. in \cite{r2m2}, that a mass matrix of
the form (\ref{mfact}), with $\balpha$ rotating with changing scale, will 
automatically give rise to the following effects: 

\begin{itemize}

\item {\bf (D1)} Mixing between the up and down states, i.e. a nontrivial CKM 
matrix between up and down quarks, and a PMNS matrix between the charged leptons
and neutrinos leading to neutrino oscillations.  [This can be easily seen in, 
for example, the CKM matrix element $V_{tb}$ which is the dot product between 
the state vectors $\bt$ and $\bb$ of respectively $t$ and $b$ in generations 
space.  For (\ref{mfact}), $\bt$ is the value of $\balpha$ at the scale 
$\mu = m_t$ while $\bb$ is the value of $\balpha$ at $\mu = m_b$, and since 
$m_t > m_b$ and $\balpha$ rotates, it follows that $\bt$ and $\bb$ are not 
aligned and $V_{tb} \neq 1$, or that there is mixing.]

\item {\bf (D2)} Fermion mass hierarchy in each species, with the mass in the 
heaviest generation in (\ref{mfact}) of each species (e.g. $t$)``leaking'' to 
the lower ones (e.g. $c$ and $u$), giving each a small but nonzero mass.  
[This can be seen as follows. The state vector $\bt$ for $t$ is the vector 
$\balpha$ at $\mu = m_t$, and the state vector $\bc$ is a vector orthogonal to 
$\bt$, and having a zero eigenvalue for (\ref{mfact}) at $\mu = m_t$.  But this 
is not the mass $m_c$ for $c$, which is to be measured at $\mu = m_c$, where 
$\balpha$ will have rotated already to a different direction with nonzero 
component in $\bc$, and hence $m_c \neq 0$.]

\item {\bf (D3)} A solution to the strong-CP problem by transforming away a 
nonzero theta-angle in the QCD action by turning it, via rotation, into a 
nonzero Kobayashi-Maskawa phase in, and giving CP-violation to, the CKM matrix. 
[At every $\mu$, the mass matrix (\ref{mfact}) has 2 zero eigenvalues, so that a
chiral transformtion can be performed to eliminate the theta-angle from the
QCD action without making the mass matrix complex.  The effects of this chiral
transformation, however, is transwmitted by rotation to other $\mu$ values and
make the CKM matrix complex leading to a KM phase.]

\end{itemize}

Any R2M2 scheme with a rotating rank-one mass matrix will give {\bf (D1) - (D3)}
but the details will depend on the rotation trajectory of $\balpha$, i.e how it
actually changes with scale $\mu$.  Let us see now what FSM has to say about
this trajectory.  Recall that $\balpha$ is itself a global quantity with no
gauge interactions, and therefore not subject directly to radiative corrections.
But it is coupled to the strong vacuum, and if that rotates with scale $\mu$, 
$\balpha$ will rotate also.  Now, information of how the strong vacuum rotates
can be obtained by studying the renormalization of any quantity which depends on
the strong vacuum.  We have, mainly for historical reasons, focussed on the
Yukawa coupling of the strong framon, and obtained therefrom the renormalization
group equations; hence also the equations governing the rotation of $\balpha$.
The implications of the rotation equation can be divided conveniently into 2
bits:

\begin{itemize}

\item The shape of the curve $\Gamma$ traced out by $\balpha$ on the unit
sphere in generation space,

\item The variable speed with respect to scale $\mu$ at which this curve
$\Gamma$ is traced.

\end{itemize}  

\begin{figure}
\centering
\includegraphics[height=17cm]{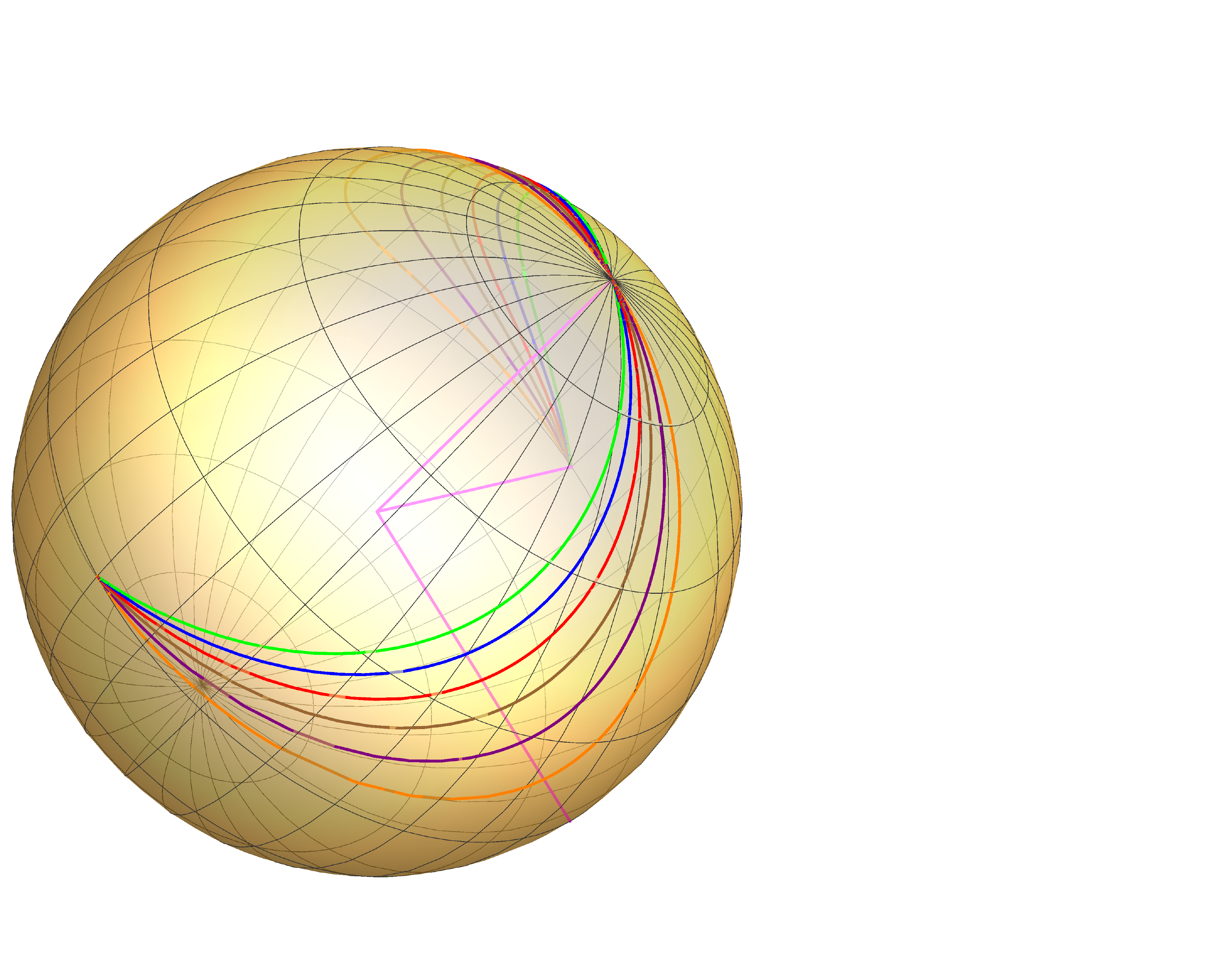}
\caption{The curve $\Gamma$ traced out by the rotating $\balpha$ on the unit
sphere in generation space, for various values of the integration constant $a$}
\label{Mui1}
\end{figure}

The shape of the curve $\Gamma$ turns out to be a consequence just of symmetry
risidual in the problem and depends only on a single integration constant, say
$a$.  This is shown in Figure \ref{Mui1}, where it is seen that $\Gamma$ bends 
sharply near $\theta - 0 , \phi = \pi$, thus giving it there a considerable 
local value for the geodesic curvature $\kappa_g$, especially for a small value 
of $a$.  But $\kappa_g$ needs not be large elsewhere, and indeed changes sign 
further along the curve.  [Notice that for a $\Gamma$ on the unit sphere, the
torsion $\tau_g = 0$ and the normal curvature $\kappa_n = 1$, and only the
geodesic curvature $\kappa_g$ is variable.]

The shape of $\Gamma$ in Figure \ref{Mui1} then immediately implies:

\begin{itemize}

\item {\bf (E1)} The corner elements, $V_{ub}, V_{td}$ of the CKM matrix for 
quarks, and similarly $U_{e3}, U_{\tau1}$ of the PMNS matrix for leptons, both 
due to twist in $\Gamma$, are much smaller than the other off-diagonal elements
because $\tau_g = 0$.

\item {\bf(E2)}  The elements $V_{us}, V_{cd}$ in the CKM matrix for quarks, due
to sideways bending of $\Gamma$ (i.e. governed by $\kappa_g$) can be much larger
than the elements $V_{cb}, V_{ts}$ governed by $\kappa_n = 1$, although the 
corresponding elements in the PMNS matrix can all be of similar magnitudes.

\end{itemize}
and also the following, though less obviously, from the already noted fact that
$\kappa_g$ eventually changes sign:

\begin{itemize}

\item {\bf (E3)} $m_u < m_d$, despite that for the two heavier generations,
$m_t \gg m_b, m_c \gg  m_s$.

\end{itemize}

The points {\bf (E1)} and {\bf(E2)} are seen to be borne out by experiment
which give the approximate CKM and PMNS matrices as:
\begin{equation}
V_{CKM} = \left( \begin{array}{ccc} 0.97428 & 0.2253 & 0.00347 \\
                                   0.2252 & 0.97345 & 0.0410 \\
                                   0.00862 & 0.0403 & 0.999152 
                \end{array} \right); \ \ \ 
U_{PMNS} = \left( \begin{array}{ccc} 0.82 & 0.55 & 0.17 \\
                                    0.50 & 0.52 & 0.70 \\
                                    0.30 & 0.66 & 0.70 \end{array} \right).
\label{CKMPMNS}
\end{equation}
The last {\bf(E3)} is of course a crucial empirical fact, without which the 
proton would be unstable, and we ourselves would not be here, but a fact which 
is, at first sight, theoretically very hard to understand.

Apart from other things, the equation governing the speed at which $\Gamma$ is
traced shows that $\balpha$ has a fixed point at $\mu = \infty$, so that its
rotation starts slowly as $\mu$ lowers from $\infty$, but will accelerate with 
decreasing $\mu$.  This then immediately implies the following results:

\begin{itemize}

\item {\bf (F1)} Since lower generation masses according to {\bf (D)(ii)} comes
from ``leakage'' through rotation and will increase with rotation speed, it 
follows from the fact that $m_t > m_b > m_\tau$ that:
\begin{equation}
m_c/m_t < m_s/m_b < m_\mu/m_\tau.
\end{equation}
This agrees with experiment which obtained the mass ratios as respectively
0.0074, 0.0227, 0.0595.

\item {\bf (F2)} Since mixing, according to {\bf (D)(i)}, also comes from 
rotation and will increase with rotation speed, it follows from the fact that 
quarks are generally heavier than leptons that the off-diagonal mixing elements 
in the CKM matrix for quarks will generally be larger than the correspondent 
elements in the PMNS matrix for leptons.  That this also agrees with experiment 
can be seen in (\ref{CKMPMNS}).

\item {\bf (F3)} Since the rotation is generally slow at quark mass scales, one 
can make a small angle approximation, which allows one to estimate the amount of
CP-violation in the CKM matrix obtained via rotation as per {\bf (D3)} from any 
given value the theta-angle in the QCD action, giving a Jarlskog invariant as
\cite{r2m2}:

\begin{equation}
|J| \sim 7 \times \sin (\theta/2) 10^{-5},
\end{equation}
which, for $\theta$ of order unity, is of the same order of magnitude as the 
experimentally measured value of $J \sim 2.95 \times 10^{-5}$.

\end{itemize}

Since, however, one has already derived the renormalization group equations
governing the rotation for $\balpha$, there is no need at all to stop at just
this qualitative level.  True, the rotation equations still depend at present
on a number of parameters, but if one is willing to supply some experimental 
information to determine these parameters, then one can proceed to evaluate 
essentially all the masses and mixing angles which are taken as inputs from 
experiment in the usual formulation of the standard model.  What the FSM does 
essentially is to replace 17 independent parameters of the standard model by 7.

\begin{itemize}

\item {\bf(G)} By fitting the 7 adjustable parameters of the rotation equations
in FSM to 6 pieces of data, we have then calculated the values of 17 independent
parameters of the standard model.  Not all the 17 quantities have been measured 
experimentally, but of those 12 that have been measured, the agreement is good.
\cite{tfsm} 

\end{itemize}
I shall leave you to judge for yourselves the significance of this result which 
you will hear from my collaborator Tsou Sheung Tsun in the next talk (II).  But
it does seem that, with the freedom still left in theory, there will not be much
difficulty in reproducing the data as are known at present.

However, what proves a theory, of course, is not its ability to reproduce known 
results, but its predictions which can be tested against and then confirmed by 
experiment.  Here, one is unfortunately hampered by not knowing how to calculate
in general with a theory where the mass matrix rotates. To deduce the results 
reported, we had some patchy rules for calculating single particle properties 
like masses and mixing angles, but we have not worked out logically the rules 
for more general calculations, so any predictions we can make at this stage are 
only tentative.  

Two tentative predictions, however, stand out, both concerned with the Higgs 
boson.  The rotating mass matrix for quarks and leptons were deduced, from a 
Yukawa coupling.  So, if $\balpha$ in the mass matrix rotates, it would seem 
that it should do the same in the Yukawa coupling too, and so affect the decay 
of the Higgs boson to fermion-antifermion pairs.  An analysis along these lines 
then suggests that:

\begin{itemize}

\item {\bf (H1?)} Branching ratios of the Higgs boson to the second generation 
fermions such as $H \rightarrow \mu^+ \mu^-, c \bar{c}$ would be suppressed 
compared to the standard model predictions.

\item {\bf (H2?)} There can be flavour-violating decays, such as $H \rightarrow 
\tau \mu$ with branching ratios at the $10^{-4}$ level.

\end{itemize}
Experimentally, present sensitivity for these decays are still a couple of 
orders of magnitude higher than is required for these effects to be detected.

The other tentative prediction involve the mixing between the weak framon (Higgs
boson) and the strong.  Given that the FSM has to be invariant under the local
gauge symmetry $G$ and also under its dual the global symetry $\tilde{G}$, the
form of the interaction potential between the framons can be worked out, which
has already been used above in the calculations reported.  It is then an easy 
step to evaluate the mass matrix of the framons to tree-level, and this shows:

\begin{itemize}

\item {\bf (H3?)} There is mixing between the Higgs boson with several of its 
strong analogues.  This mixing depends on a couple of yet unknown parameters and
so its details are still unknown, but could make the Higgs' decays depart from 
the standard model predictions.

\end{itemize}

The predictions {\bf (H1?)-(H3?)} all come from the weak framon (\ref{wframon}).
What is likely to be even more interesting, however, is how the strong framons 
of (\ref{sframon}) will manifest themselves.  After all, they are the truely 
new ingredients added by the FSM to the standard model.  We recall that it is
strong framons loops which gave rise, via renormalization, to rotation in the 
mass matrix of quarks and leptons, and hence to their mixing and mass hierarchy.
So these important effects should already be regarded, in the FSM context, as 
indirect manifestations of the strong framon.  Thus, rather, the question that 
one is actually asking now is whether strong framons could manifest themselves 
more directly in some other experimentally detectable physical phenomena.  The
answer to this, however, is not obvious and is at present perforce speculative,
the strong framon being an entirely new type of field with special properties,
including probably some unusual soft (nonperturbative) colour physics.  But the
question is nevertheless sufficiently intriguing, with potentially far-reaching 
consequences, to deserve some speculations, to which let us then, for a little 
while, indulge.

We recall that the strong framon carries colour, and so, because of colour 
confinement, cannot exist as a particle in free space, only inside hadronic 
matter.  But it can combine with other consituents of hadronic matter with the
opposite colour to form a colour neutral state, which then appear as a hadron
in free space.  For example, a framon can combine with an antiframon to form
bosons, which in the lowest s-wave state are colour analogues of the standard
Higgs boson in electroweak $su(2)$ and mix with it, as mentioned in {\bf (H3?)}
above.  

There are altogether 9 such states, and they are likely to be the lowest batch 
of the new hadrons which contain a strong framon as constituent.  As also 
mentioned, their mass matrix at tree-level has already been worked out from the
known framon self-interaction potential, and it is straightforward to extend the
calculation to their mutual couplings, which has now also been completed.  These
show that they can readily decay into one another, but the lowest among them,
called, for historical reason, $H_-, H_4, H_5$ are, of course, stable against 
such decays.  These $H_K$'s are electrically neutral, but have for constitutents
the strong framons (\ref{sframon}) with charge $-1/3$, which would allow, among 
them, $H_-$, made from a framon and an antiframon with the same generation
$\widetilde{su}(3)$ index, to decay into photons.  But this does not apply to
$H_4$ and $H_5$ which are made from a framon and antiframon carrying different
$\widetilde{su}(3)$ indices.  Further, since even their framon constituents 
carry no weak charge, these last $H_f$'s ($f$ = four or five) also cannot decay
weakly, and would thus seem to be stable altogther.

Now, if these $H_f$ states do exist and are stable, an obvious question would be
why they have not been seen.  Presumably, like quarks and gluons, strong framons
would be present in the sea of hadronic matter inside a proton.  And since they
are both coloured and charged, they too, like quarks, can be ``knocked out'' by
a hard kick from a gluon or a photon.  A quark so ``knocked out'' will pick up 
and combine with an antiquark in the sea and emerge as a meson.  Cannot then a
``knocked out'' framon pick up and combine with an anitframon from the sea and
emerge as a $H_f$?  If so, why do we not see it?

There is a difference, however, between a quark and a framon in that the latter
has an imaginary mass, for like the Higgs scalar field, the quadratic term in 
the self-interaction potential has a negative coefficient.  One can interpret
this as meaning that the framon in hadronic matter, unlike the quark, has only 
a finite life-time.  It has thus only a limited time to seek out a partner from
the sea to form an $H_f$, and in this it may not succeed if the time is short. 
Hence, again unlike the quark, a ``knocked out'' framon may not succeed to
emerge from the host proton at all.  Whether this happens will depend on its
life-time and the amount of hadronic matter that it can traverse inside the 
proton during its life-time.  Let us say here, for the sake of argument, that 
the conditons are such that this will not happen, so that no $H_f$'s can be
produced in ordinary hadronic reactions, and so explain their non-observation so
far in experiment. 

However, that no $H_f$'s are produced in ordinary hadron collisions in present 
experiments need not mean that the same is true under other circumstances.  For
example, in the primordial universe (or even now in, say, the galactic centre)
both temperatures and densities are much higher than can be found under present
experimental conditions in our laboratories.  This may allow then these $H_f$
to be formed, and once formed, being stable according to our previous argument,
they would still be around with us today.

The question then leaps out whether they may be candidates for dark matter.  A
preliminary investigation does indeed indicate that they may have rather little
interaction with ordinary matter, and also with themselves.  That the strong
framon is short-lived suggests, by an argument similar to that given above for
the non-production of $H_f$'s in present laboratory experiments, that they do
not have the usual strong interactions of ordianry hadrons (although they are
formally hadrons, being colour neutral bound states by colour confinement of 
coloured constituents). Besides, they are charged neutral, both electrically
and in colour.  And being relatively light (though not light enough to be hot 
dark matter) it may not be impossible for them to satisfy the otherwise very 
stringent bounds already set by recent experiments such as Xenon 100 and LUX.  
Hence, perhaps:

\begin{itemize}

\item {\bf (I??)} New constituents of Dark Matter (?)  

\end{itemize}

However, if indeed relatively light, it will take a lot of them to make up a
sufficient mass so as to matter in the dark matter problem.  Is there then any 
reason why they should be produced in the early universe in such an abundance,
say, as compared to luminous baryonic matter?  Amusingly, there is, or at least 
may be, a possible reason.  At some stage in its development after the Big Bang,
the universe is presumably just a large blob of hot, dense hadronic matter with,
among other things, quarks and framons swimming around.  As it cools and expands
further, the quarks and framons inside would be frantically seeking partners to
hitch up as colour neutral bound states so as to survive into the next epoch as 
hadrons.  For the framon to survive as an $H_f$, all it needs is to find an
antiframon of opposite colour.  For the quark to survive as a nucleon, however,
it will have to find 2 others of the right colour at the same time, which looks
an altogether tougher proposition.  Hence, the much greater abundance of $H_f$
than nucleons in our world today.  Indeed, it would seem a very lucky chance
that enough luminous baryonic matter managed to survive, or else most of the
things we know would not be here.

As matters stand, of course, all this discussion about the $H_f$'s as dark 
matter is merely an exercise in imagination.  But, since some of the parameters
in FSM have already been determined in the work to be reported by Tsou in the 
next talk, there may be a chance that some of these imaginings can actually be 
investigated, and either substantiated or repudiated in the not too distant 
future.

If there is any truth in the these speculations, however, it would indeed seem 
an extraordinary stroke of good luck that enough luminous baryonic matter is 
left around from the early universe for us humans to come into existence.  Then
we are even luckier than we think, today, to be able to come together here to
celebrate the 60th anniversary of the Yang-Mills theory.  For this we have to 
thank, first Professor Yang for giving us the cause to celebrate, and secondly 
Professor Phua and the other organisers for giving us the opportunity to enjoy 
this celebration.

The work summarized in this talk has almost all been done in collaboration with
Jose Bordes, and in part in collaboration with Mike Baker.  We have benefitted
also from discussions with, and constant interest and encouragement from,
James Bjorken.

\end{document}